\documentclass[useAMS,usenatbib,usegraphicx]{mn2e}
\usepackage{epsfig}
\usepackage{amsmath} %need for line break in equ 3

\usepackage{times}

\def\kms{km ${\rm s}^{-1}$}

\def\ch2{$\chi^2$}

\def\kms {\hbox{${\rm km\ s}^{-1}$}}

\def\scm  {$\hbox{{\rm cm}}^{-2}$}    %cm-2
\def\MOLH {\hbox{${\rm H}_2$}}  %H2
\def\rr {\rightarrow}

\def \HI {H{\sc \,i}}

\def\lapp{\ifmmode\stackrel{<}{_{\sim}}\else$\stackrel{<}{_{\sim}}$\fi}
\def\gapp{\ifmmode\stackrel{>}{_{\sim}}\else$\stackrel{>}{_{\sim}}$\fi}

\title[Redshifted Radio Absorption Lines I]{A Survey for Redshifted Molecular and Atomic Absorption Lines \\
{\Large I. The Parkes Half-Jansky Flat-Spectrum Red Quasar Sample}}
\author[S. J. Curran et al.]{S. J. Curran$^{1}$\thanks{E-mail:
sjc@phys.unsw.edu.au}, M. T. Whiting$^{1,2}$,
M. T. Murphy$^{3}$, J. K. Webb$^{1}$, S. N. Longmore$^{1,2}$,
\newauthor 
Y. M. Pihlstr\"{o}m$^{4}$, 
R. Athreya$^{5}$
and C. Blake$^{6}$\\
$^{1}$School of Physics, University of New South Wales, Sydney NSW 2052, Australia\\
$^{2}$CSIRO Australia Telescope National Facility, PO Box 76, Epping NSW 1710, Australia\\
$^{3}$Institute of Astronomy, Madingley Road, Cambridge CB3 0HA, UK\\
$^{4}$Department of Physics and Astronomy, University of New Mexico, 800 Yale Boulevard NE, Albuquerque, NM 87131, USA\\
$^{5}$National Centre for Radio Astrophysics, Pune 411 007, Maharashtra, India\\ 
$^{6}$Department of Physics \& Astronomy, University of British Columbia, 6224 Agricultural Road, Vancouver, B.C., Canada V6T 1Z1
}

\begin{document}

\date{Accepted ---. Received ---; in original form ---}

\pagerange{\pageref{firstpage}--\pageref{lastpage}} \pubyear{2006}

\maketitle

\label{firstpage}

\begin{abstract}
We are currently undertaking a large survey for redshifted atomic and
molecular absorption lines at radio frequencies. In this paper we
present the results from the first phase of this: the search for \HI\
21-cm and OH 18-cm absorption lines in the hosts of reddened quasars
and radio galaxies. Although we observed each source for up to several
hours with two of the world's most sensitive radio telescopes, the
Giant Metrewave Radio Telescope (GMRT) and Westerbork Synthesis Radio
Telescope (WSRT), only one clear and one tentative detection were
obtained: \HI\ absorption at $z = 0.097$ in PKS 1555--140 and OH
absorption at $z =0.126$ in PKS 2300--189, respectively, with the
Australia Telescope Compact Array (ATCA). For the latter, no \HI\
absorption was detected at the same redshift as the borderline OH
detection. In order to determine why no clear molecular absorption was
detected in any of the 13 sources searched, we investigate the
properties of the five redshifted systems currently known to exhibit
OH absorption. In four of these, molecules were first detected via
millimetre-wave transitions and the flat radio spectra indicate
compact background continuum sources, which may suggest a high degree
of coverage of the background source by the molecular clouds in the
absorber. Furthermore, for these systems we find a relationship
between the molecular line strength and red optical--near infrared
($V-K$) colours, thus supporting the notion that the reddening of
these sources is due to dust, which provides an environment conducive
to the formation of molecules. Upon comparison with the $V-K$ colours
of our sample, this relationship suggests that, presuming the
reddening occurs at the host galaxy redshift at least in some of the
targets, many of our observations still fall short of the sensitivity
required to detect OH absorption, although a confirmation of the
``detection'' of OH in 2300--189 could contravene this.

\end{abstract}


\section{Introduction}
\label{intro}

Unlike optical spectral lines, the \HI~21-cm spin-flip transition of
neutral hydrogen and molecular rotational transitions are transparent
to interstellar dust, thus providing very useful probes of the early
Universe. For example, these can be used to:

\begin{enumerate}

\item Investigate neutral gas at high redshift allowing a measure of the
  baryonic content of the early Universe, when neutral gas
  outweighed the stars: At low redshift, most of the gas has been
  consumed by star formation and \HI ~represents only a small
  fraction of the total mass in baryons. Conditions were very
  different at high redshift and quasar absorption line studies have
  demonstrated that the co-moving density of \HI ~is many times
  higher at $z \sim 3$ (e.g. \citealt{psm+01}).
  
\item Probe the evolution of large-scale
  structure. In the most successful models, massive galaxies build up through
  merging and accretion.  This process is most vigorous at high redshift
  where interactions occur more frequently. \HI ~observations are a
  powerful probe of the dynamics of these mergers, and thus constitute a
  basic test of theories of galaxy formation and evolution.

\item High redshift observations of \HI\ can give a lower limit to
  the epoch of reionisation, when neutral hydrogen collapsed
  to form the first structures (stars, galaxies/quasars), the upper
  limit of which is constrained by Cosmic Microwave Background (CMB)
  measurements (e.g. \citealt{gbl99}).

\item The relative populations of molecular rotational transitions can
  measure density and temperature at high redshift, thus providing a
  probe of the CMB and the abundance of cold star
  forming gas in the early Universe.
  
\item Monitoring absorbers which act as gravitational lenses can yield
  time delay studies giving measurements of the Hubble parameter (e.g. \citealt{wc01}).

\item Various combinations of \HI, OH and millimetre rotational lines
  can give accurate measurements of several fundamental constants; the
  electron--proton mass ratio the proton g-factor and the fine
  structure constant (see \citealt{cdk04} and references
  therein). These can provide at least an order of magnitude increase
  in accuracy over the current optical results, which may suggest that
  fine structure constant has undergone cosmological evolution
  (\citealt{mwf03}, although see \citealt{scpa04}).
\end{enumerate}

Unfortunately, redshifted radio absorption systems are currently very
rare, with only 50 \HI\ absorption systems known for redshifts of
$z\gapp0.1$ (summarised in Table \ref{t2}). Five of these systems also
exhibit OH absorption, four of which constitute the only known
redshifted millimetre absorption systems (see Table
\ref{t3}).
%The four OH systems are included in the number of known \HI\ absorbers and
%constitute the only known redshifted millimetre absorption systems \citep{wc94,wc95,wc96b,wc98}. All occur towards faint objects with $V\gapp20$. 
In addition to these, molecular absorption has also been detected in
10 known optical absorbers through \MOLH\ vibrational transitions
redshifted into the optical band at $z\gapp1.8$ (see
\citealt{rbql03,cbgm05}).  These occur in damped Lyman-alpha
absorption systems (DLAs), which have high neutral
hydrogen column densities ($N_{\rm HI}\geq2\times10^{20}$ cm$^{-2}$)
known to exist at precisely determined redshifts. 

\begin{figure}
\vspace{8.0 cm} \setlength{\unitlength}{1in} 
\begin{picture}(0,0)
%\put(-0.3,-0.25){\special{psfile=colour_fraction_plot.ps hscale=70 vscale=70 angle=0}}
\put(-0.2,3.7){\includegraphics{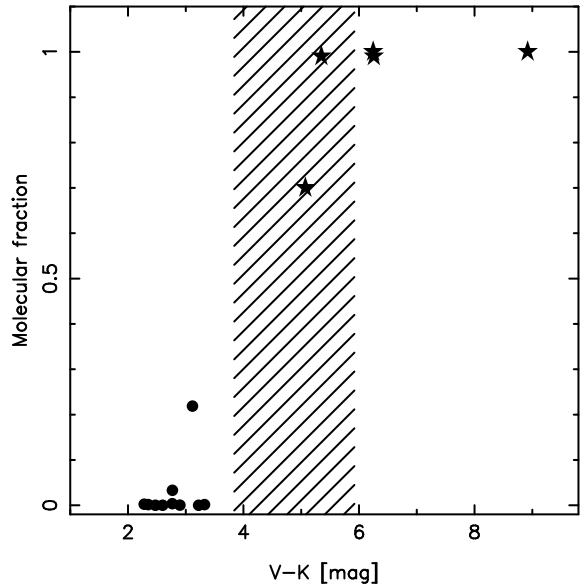}}
\end{picture}
\caption{Molecular fraction versus optical--near-infrared colour in
cosmological absorbers.  The circles represent the H$_2$-bearing DLAs
and the stars the known OH absorbers, for which two are coincident on
the plot at a molecular fraction of near unity and
$V-K\approx6.2$. The hatched region shows the $V-K$ colour range of
our sample.}
\label{f1}
\end{figure}

However, extensive millimetre-wave observations of DLAs have yet to
detect absorption from any rotational molecular transition
\citep{cmpw03}, leading us to suspect that using optically-selected
objects selects against dusty environments, which are more likely to
harbour molecules in abundance. This is apparent when one compares the
DLAs in which H$_2$ has been detected, which have molecular fractions
${\cal F}\equiv\frac{2N_{\rm H_2}}{2N_{\rm H_2}+N_{\rm
HI}}\sim10^{-7}-10^{-2}$ (see \citealt{cwmc03} and references therein)
and colours of $V-K=2.2 - 3.4$ (typical for optically-bright quasars)
with the five known OH  absorbers; ${\cal F}\approx0.7-1.0$
(using $N_{\rm H_2}\sim10^7N_{\rm OH}$, see \citealt{cw98,
kc02a,kcl+05}) and $V-K\geq5.07$ (Fig. \ref{f1}). This indicates
substantial reddening of the quasar light in these cases, possibly by
the absorber,
which is either intrinsic to the source (in the cases of 1504+377 and
1413+135) or due to the lensing galaxy (for 0132--097, 0218+357 and
1830--211). We have therefore commenced a survey for redshifted atomic and
molecular absorption lines towards reddened
objects. In this paper we present the results of the first phase of
this survey: searching for \HI\ and OH lines due to cold dense dust residing in the
hosts of quasars and radio galaxies.

%and at such high H{\sc i} column densities molecular fractions of$f({\rm H}_{2})\sim1$ may be expected \citep{sch01}.

\section{Observations}
\label{observations}

\begin{table*} 
\centering
\begin{minipage}{175mm}
\caption{The targets searched and the observational
results. $^*$Denotes the radio galaxies (see main text). $\nu_{\rm
obs}$ is the observed frequency range of the line [MHz], $\sigma_{{\rm
rms}}$ is the r.m.s. noise [mJy] reached per $\Delta v$ channel
[\kms], $S_{\rm cont}$ is the continuum flux density, $\tau$ is the
optical depth of the line calculated per channel, where
$\tau=3\sigma_{{\rm rms}}/S_{\rm cont}$ is quoted for the
non-detections. In all cases OH refers to the $^{2}\Pi_{3/2} J = 3/2$
(1667 MHz) transition, with the exception of 1535$+$004 for which we observed the
$^{2}\Pi_{1/2} J = 1/2$ (4751 MHz) transition. Finally, we list the $K$-band magnitude and the
optical--near-infrared colours.}
\begin{tabular}{@{}l l c c c c c c c c c c  c c@{}} 
\hline
Source          &$z_{\rm host}$
                         &Line  &$\nu_{\rm obs}$ & $\sigma_{{\rm rms}}$ 
                                                    & $\Delta v$ 
                                                          & $S_{\rm cont}$ 
                                                                 & $\tau$    & Tel. & $K$    & $R-K$   & $V-K$ & Ref & Notes \\
\hline
%HAVE COMMENTED OUT THOSE YET TO BE REDONE
0003$-$066     & 0.347   & OH   & 1244 &-- & -- &--  & -- & ATCA & 14.12  & 3.69    & 3.85  & 1,2 & $a,b$\\
0108$+$388     & 0.66847 & OH   & 996--1003      & 2.9 & 19  & 0.20 & $<0.044$  & GMRT & 16.69  & 5.31    & --    & 3,4 & \\
0114$+$074$^*$ & 0.342   & OH   & 1239--1246     & 12 & 15  & 2.90 & $<0.012$  & GMRT & 15.39  & --      & 5.25  & 1,2 & $a$ \\
0131$-$001     & 0.879   & \HI  & 756  & 19  &--   & -- & --  & WSRT & 16.78  & 4.00    & 5.72  & 5   & $b$\\
 ...           &...      & OH   & 880--897  & 6.2 & 6.6   & 0.94 & $<0.020$  & WSRT & ...    & ...     & ...   & ... & $b$\\
%%0132--097    & 2.220   & OH 6-cm & 1447        & 27  & 13  & 1.69 & $<0.049$  & GMRT & 13.55  &         & 8.92  &     &  \\
0153$-$410$^*$ & 0.226   & OH   & 1353--1378     & 34 & 28  & 1.87 & $<0.05$  & ATCA & 14.14  & 3.27    & 4.01  & 5  & \\
%%%0202+149    &0.405    & --   & 1238--1246     & 3.7 & 15  & 1.00 & $<0.011$  & GMRT & 16.51  & 4.13    & 5.82  &     &  \\
0202$+$149     & 0.405   & OH   & 1183--1190     & 5.0 & 15  & 2.20 & $<0.007$  & GMRT & 16.51  & 4.13    & 5.82  & 6   & \\
0454$+$066     & 0.405   & \HI  & 1007--1012     & --  & 18  & --   & --        & GMRT & 14.424 & 3.669   & 4.318 & 7   & $c$\\
0500$+$019     & 0.58457 & OH   & 1049--1056     & 7.4 & 18  & 1.75 & $<0.013$  & GMRT & 15.430 & 5.252   & 5.916 & 1,6 & \\
1107$-$187     & 0.497   & \HI  & 948.8          & --  & --  & --   & --        & GMRT & 15.96  & 3.34    & 5.14  & 5   & $d$\\
..             & ...     & OH   & 1110--1117     & 2.4 & 17  & 0.63 & $<0.011$  & ...  & ...    & ...     & ...   & ... & ...\\
1353$-$341$^*$ & 0.2227  & OH   & 1358--1384     & 9.7 & 27  & 0.57 & $<0.051$  & ATCA & 13.62  & 3.12    & 3.84  & 5   & \\
1355$+$441     & 0.6451  & OH   & 1010--1017     & 1.9 & 18  & 0.65 & $<0.0088$ & GMRT & $>16.17$ & $>4.78$ & --  & 2,8 & \\
1450$-$338     & 0.368   & \HI  & 1035--1042     & 10 & 18  & 1.49 & $<0.020$  & GMRT & 15.23  & 4.16    & 5.17  & 5   & \\
...            & ...     & OH   & 1212--1238     & 9.3 & 31  & 1.07 & $<0.026$  & ATCA & ...    & ...     & ...   & ... & \\
1535$+$004     &  3.497  & \HI  & 315.0--316.6   & 7.3 & 15  & 0.36 & $<0.061$  & GMRT & 19.54  & $>2.76$ & --    & 1,5 & \\
... & ... &OH & 1053--1060  & 6.1 & 18  & 0.57 & $<0.032$  & GMRT & ...    & ...     & ...   &  ...   & \\
1555$-$140$^*$ & 0.097   & \HI  & 1289--1312     & 3.7 & 29  & 0.31 & 0.031     & ATCA & 13.08  & 3.20    & 3.85  & 5   & \\ 
2300$-$189$^*$ & 0.129   & \HI  & 1255--1277     & 12 & 30  & 1.21 & $<0.029$  & ATCA & 13.05  & 3.51    & 3.88  & 1   & $a$ \\ 
..             & ...     & OH   & 1470--1496     & 7.2 & 25  & 1.04 & $<0.021$ & ATCA & ...    & ...     & ...   & ... & \\
2337$-$334     & 1.802   & OH   & 593--597       & 9.2 & 16  & 2.61 & $<0.011$  & GMRT & 16.39  & 5.01    & 5.50  & 5   & \\
%%2252--090    & 0.6064  & OH   & 1034--1042     & --  & 18  &\multicolumn{2}{c}{WRONG COORDS} & GMRT &21.5 & 5.0 & \\
\hline 
\end{tabular}
{\flushleft Photometry references: (1) Francis [priv.\ comm.], (2)
\citet{hmr+01}, (3) \citet{srkr96}, (4) \citet{sobl93}, (5)
\citet{fww00}, (6) This paper [see Sect. \ref{sec-phot}], (7)
\citet{whi00}, (8) 2MASS.\\ Notes: $^a$$V-K$ estimated using
$B-V\approx0.5$, the mean value for a PHFS quasar, and $B-V\approx1.5$
the mean value for a PHFS radio galaxy, $^b$RFI in some channels
required splicing of the observed band, $^c$RFI prevents a confident
limit, $^d$RFI so severe that fringes could not be obtained for the
bandpass calibrator.}
\label{obs}
\end{minipage}
\end{table*} 

Details of our targets are given in Table \ref{obs}. For the most part
they have been selected from the Parkes Half-Jansky Flat-spectrum
Sample (PHFS, \citealt{dwf+97}) on the basis of their optical--near-IR
photometry. The nature of the PHFS (bright, generally compact radio
sources), together with the excellent optical information
\citep{fww00} and the near completeness of the sample, makes it an
ideal data base from which to select sources. Our targets
are amongst the reddest objects in the PHFS, and many have colours
believed to be due to dust extinction \citep{wfp+95}.  Note that in
addition to the PHFS sources, we include 0108+388, 0500+019 and
1355+441 in the search for OH absorption, since these are reddened (or
at least optically faint) quasars already known to exhibit H{\sc i}
absorption at the host redshift (see Table \ref{sum} for details).

\subsection{Optical/Near-Infrared photometry}
\label{sec-phot}
%MATT - WHERE THESE COME FROM (COLUMN IN TABLE) AND WHAT WE REOBSERVED\\

As stated above, the key requirement in our object selection is a red
optical--near-infrared colour, and we have therefore selected objects
with $V-K\gapp4$. Our selection is of course subject to the frequency ranges
accessible by the radio telescopes used in our survey and so not all
of the PHFS sources with $V-K\gapp4$ are included.

In addition to the colour requirement, in order to broaden our sample,
we also select several radio galaxies from the PHFS whose optical
spectra \citep{dwf+97} exhibit a lack of broad lines such as
H$\alpha$: These ``type-2'' objects may have the line-of-sight to
their nuclear regions obscured by the putative ``dusty torus''
\citep{ant93}, thus providing a reservoir of molecular gas. Since
their integrated photometry is dominated by starlight, the
optical--near-infrared colours of these objects show colours typical
of elliptical galaxies. The radio galaxies searched are indicated in
Table~\ref{obs}.

In the table we also give the $K$-magnitude ($2.2~\mu$m) and two
estimates of colour; $R-K$ and $V-K$. Most of these sources are part
of the large dataset of \citet{fww00}, which provides all of the
necessary optical and near-infrared photometry. A further four sources
use $K$ photometry from the Anglo-Australian Telescope's {\it IRIS}
instrument, taken during the compilation of the PHFS (Francis, priv.\
comm.).  For these sources, the $V$ magnitude is estimated from the PHFS
$B$ magnitude by assuming $B-V=0.5$ (for 0003$-$066) and $B-V=1.5$ (for
0114$+$074 \& 2300$-$189), see Table \ref{obs}. The $R$ magnitudes for
these sources were taken from the SuperCOSMOS Sky Survey
\citep{hmr+01}.  Both the optical and near-infrared photometry for
0454$+$066 was taken from \citet{whi00}.

Two sources (0202$+$149 \& 0500$+$019) were observed in the optical
(and the near-infrared in the case of 0202$+$149) at the ANU 2.3-metre
telescope over two runs, in August 2004 and April 2005, respectively,
at the same time as other projects. The data were reduced with
standard techniques (the method was very similar to that described in
\citealt{fww00}). Optical photometry in $BVRI$ bands was obtained with
reference to Landolt standard fields \citep{lan92}, using the IRAF
{\sc photcal} package, while the near-IR photometry was referenced to
IRIS standards \citep{mhd+94}.

Finally, both 0108$+$388 and 1355$+$441 are too far north to be
observed from Siding Spring.  For 0108$+$388 we use photometry of
\citet{srkr96} [$K$-band] and \citet{sobl93} [$R$-band]\footnote{There
is no published $V$-band measurement and it is too faint to appear on
SuperCOSMOS}. 1355$+$441 is marginally detected on SuperCOSMOS at
$R$-band, but not detected in 2MASS, so we quote the $3\sigma$ upper
limit to $K$ in Table~\ref{obs}.

\subsection{ATCA observations}
\label{atca}
As seen from Table \ref{obs}, the ATCA sources are generally not as
red as the rest of the sample, with $V-K\approx4$ for four of the
five sources. As mentioned previously, the
sources selected ultimately came down to the radio band available at
the telescope and so it was not always possible to select sources
based purely on their colour. We therefore included these moderately
red sources which could be observed with the ATCA, since most of them
are radio galaxies (Sect. \ref{sec-phot}) and they may prove useful in
comparing any established connection between the \HI, OH abundances
and the degree of reddening of the source.

The observations were performed over 2 nights in July 2004. We used the H168 array,
giving a maximum baseline of  4469 metres. For the front-end we used 
the H--OH receiver (1.25 to 1.78 GHz) backed by the FULL\_32\_256
correlator configuration, giving 32 MHz over 256 channels, or a
coverage of 6500--8000 \kms ~at a channel width of $\approx30$ \kms
~for the frequencies observed. For each source we tuned to the host
redshift offset by $\approx+6$ MHz in order to cover a larger
blueshift range with respect to the source, where foreground
absorption in the host may be expected. For bandpass and flux
calibration we used both PKS 0023--26 and PKS 1934--63, with self
calibration used for the antenna gains in all cases except PKS 0153--410 and
PKS 2300--189, for which we used PKS 0201--440 and PKS 2252--089,
respectively. As with the other radio data, these were reduced using the {\sc
miriad} interferometry reduction package, with which we extracted a
summed spectrum from the emission region of the continuum maps
(Fig. \ref{spectra}) and all of the
sources were unresolved by the $\gapp2'$ synthesised beam. 
Specifically for each source:\\ 
0003--066 was observed
for 3 hours, although due to severe RFI in the bandpass calibrator at
1247 MHz, the band was heavily split and no image could be
produced.\\ 
0153--410 was observed for 0.6 hours and only flagging of
some edge channels was required.\\ 
1353--341 was observed for 1.0 hour,
and again, only flagging of some edge channels was required. Although
the dip apparent towards the upper end of the band
(Fig. \ref{spectra}) appears in both polarisations (albeit
unequally), the feature appears only in a single antenna pair
(CA01--CA05) and so we consider this to be an artifact.\\
1450--338 was observed for 1.0 hour and severe RFI rendered all
antenna pairs, except those with the $>4$ km distant CA06 antenna, unusable.
For the remaining 5 antenna pairs, the data were excellent with no
flagging required, although it was not possible to produce an image with 
these limited data.\\
1555--140 was observed for 3 hours and no major flagging
of data was required. Even before extracting a spectrum from the cube 
(as shown in Fig. \ref{spectra}), the absorption line was 
apparent in both polarisations upon the averaging of the visibilities.\\
1.5 hours were spent searching for \HI\ at the redshift of 2300--189. However
0.5 hours of bad data had to be removed from the beginning of the observations
for all antenna pairs not involving CA06. In addition to this, baseline CA02--CA03
had to be removed. For OH, we observed for 6.5 hours on source and no major flagging of data was required.

%\subsection{GBT observations}

%Glen's stuff in here

\subsection{GMRT observations}
\label{gmrtobs}

The GMRT observations were performed in March \& April 2004 and
February 2005. For all runs we used all 30 antennas and the 21-cm \&
50-cm receivers over 8 \& 4 MHz bandwidths, respectively. This gave a
band of 62 kHz for the low redshift (21-cm band) sources and 31 kHz
for the high redshift (50-cm band) sources in each of the 128 channels
(2 polarisations). Like the ATCA observations, this resulted in a
fairly poor velocity resolution (Table \ref{obs}), although this was
necessary in order to cover any possible offsets between the
absorption and the host redshifts ($\lapp1500$ \kms,
\citealt{vpt+03}), as well as the large \HI\ ($\lapp500$ \kms) and OH
($\approx200$ \kms) line-widths observed in the known systems. Again,
all of the sources were unresolved by the synthesised beam, which was
typically $\gapp5''$ for these observations. For all of the runs we
used 3C48, 3C147 and 3C286 for bandpass calibration and used separate
phase calibrators for all of the sources, as heavy flagging of the
target data could result in poor self calibration. In general,
however, RFI was not too severe with only two sources being completely
lost (see below). For all runs we noted a spike in channel
117 of the RR polarisation and this was removed along with the
telescope pairing between antennas E02 and E03 (15 and 16 in {\sc
aips}/{\sc miriad} convention), which was invariably bad. In particular for each source:\\ 
0108+388 was observed for a total of
3.6 hours, with 330 good antenna pairs remaining after removing
the non-functioning antennas.\\
0114+074 was observed for 4.1 hours, however, due to RFI,
only 340 antenna pairs were used and the last 2 hours of the observation
had to be removed.\\
0202+149 was observed for 6.7 hours and minimal RFI allowed
us to retain 400 good antenna pairs.\\
Close to 1011 MHz, where 0454+066 was observed, RFI was so severe
that no reliable calibration could be obtained.\\
0500+019 was observed for 1.0 hour and minimal RFI allowed
us to retain 330 good antenna pairs.\\
1107--187 was observed at $\approx1114$ MHz, in the search for OH, for 6.9 hours with 400 good antenna
pairs being retained. The \HI\ transition fell into the same mobile phone band
which marred other observations in this band (see Sect. \ref{other}),
thus preventing even good data for the bandpass calibrator.\\
1355+441 was observed for 5.5 hours and 420 good antenna pairs were
retained.\\ 1450--388 was observed for 0.9 hours with 360 good antenna
pairs. However, due to poor gains, the first 35 minutes of the
observations had to be removed.\\
In the 90-cm band 1535+004 was observed for 1.4 hours and 390 good antenna
pairs were retained. Since the OH $^{2}\Pi_{1/2} J = 1/2$ (4751 MHz) transition
at the redshift of this source fell into the 20-cm band of the GMRT, we also observed this for 3.1 hours with the 360 good antenna pairs available.\\
2337--334 was observed for 3.8 hours and after flagging 300 antenna pairs were
retained.\\

%Three hours were also allocated for \HI\ in these sources,although severe RFI at 949 MHz, due to local mobile phone transmitters, prevented us even obtaining fringes in the bandpass calibrator. 
%The data were reduced independently by two of the authors using both {\sc aips} and {\sc miriad}.

%WHAT HAPPENED TO \HI\ IN  0202+149 and OH IN 1430-155 (HI WAS OUT OF BAND)?
%ran out of time for 0202+149
%The 552.04 MHz line towards B1430-155 lies well outside the measured bandwidth of the 610 MHz system

\subsection{WSRT observations}
\label{wsrt}

PKS 0131--001 was observed with the WSRT during a
survey for radio absorption lines in gravitational lenses (Sect. \ref{other}), using all 14 antennas and the UHF-high band receiver with 20
MHz bandwidth over 1024 channels in both polarisations. \HI ~was
searched for at a frequency of 756 MHz ($z_{\rm host}=0.879$) over
$5\times0.85 + 1\times0.26$ hour slots on 23rd--24th November
2004. Due to severe RFI from a German TV station, this frequency was
reobserved for 4 hours in $4\times0.85 + 1\times0.59$ hour slots on
19th February 2005. Again some RFI was present, particularly in one of
the polarisations (XX), which was subsequently removed. As per the
other sources observed, we also flagged channels 900--1024 ($>763.3$
MHz here). After flagging the worst frequencies (751.9--753.1 MHz,
754.3--755.4 MHz and 759.5--761.5 MHz), the baselines in which the
r.m.s. exceeded 1 Jy ($\gapp60$\% of the baselines) were also flagged.
However, as seen from the oscillations around the flagged interference at 
$\approx755$ MHz, the effects of RFI are still present and since
the expected absorption frequency of $\approx756$ MHz is coincident
with these, we cannot assign confident limits to the \HI\ optical depth.

OH 18-cm ~was searched for at a frequency of 887 MHz over
$14\times0.85$ hour slots on 17th November 2005. Although the RFI was
not as bad as at the lower frequency, five of these slots had to be
removed, as well as at least 10 channels to either side of 885.1 MHz
in all of the data.
Unfortunately, the low declination of the source coupled with the fact
that the WSRT is an E--W array, meant that the u--v coverage was
insufficient to produce any images. Therefore the limits quoted for
0131-001 in Tables \ref{obs} and \ref{sum} are from the averaged
visibilities.

%\clearpage
\section{Results and Discussion}
\subsection{Observational results}

\label{results}

\begin{figure*}
	\vspace{23cm} \setlength{\unitlength}{1in}

%\special{psfile=spectra/0003-066.ps hoffset=-20 voffset=670 hscale=45 vscale=45 angle=-90}

\includegraphics{0108+388.ps}
\includegraphics{0114+074.ps}
\includegraphics{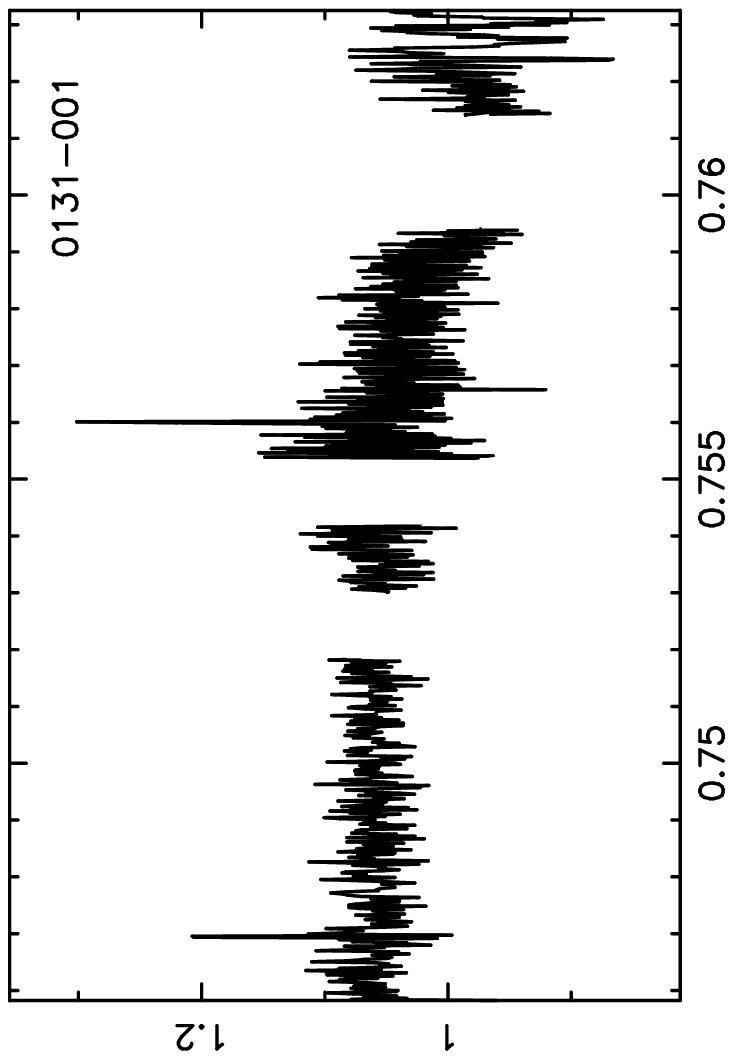}
%new line
%\special{psfile=0131-001-OH-LINE.ps hoffset=-20 voffset=540 hscale=45 vscale=45 angle=-90} %HAVE HAD TO XFIG LINE OUT
\includegraphics{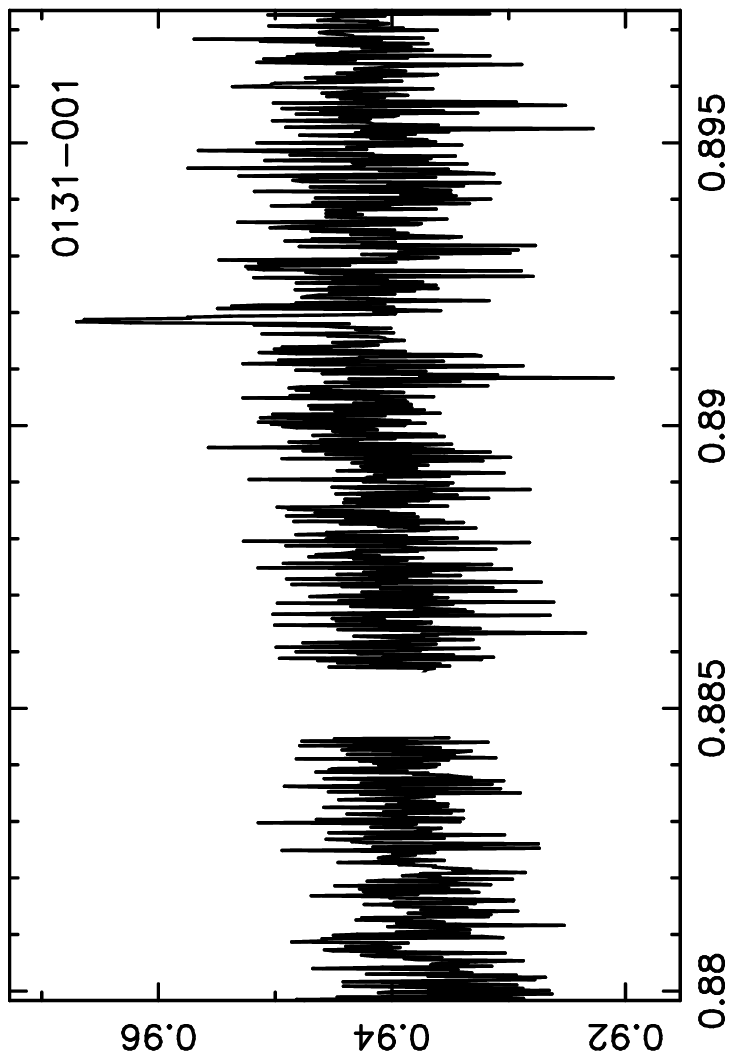}
\includegraphics{0153-410.ps}
\includegraphics{0202+149.ps}

%new line
\includegraphics{0500+019.ps}
\includegraphics{1107-187-OH.ps}
\includegraphics{1353-341.ps}
%new line
\includegraphics{1355+441.ps}
\includegraphics{1450-388-HI.ps}
\includegraphics{1450-338-OH.ps}
%new line
\includegraphics{1535+004-HI.ps}
\includegraphics{1535+004-OH.ps}
\includegraphics{1555-140.ps}
\caption{The observational results. All spectra have been extracted
from the spectral cube, with the exception of 0131--001 and 1450--338
at 1225 MHz where the visibilities are averaged together. In each
spectrum and in Figs. \ref{spectrum} and \ref{2300-spectrum} the
ordinate shows the flux density [Jy] and the abscissa the Doppler
corrected barycentric frequency [GHz].}
\label{spectra}
\end{figure*}
\begin{figure*}
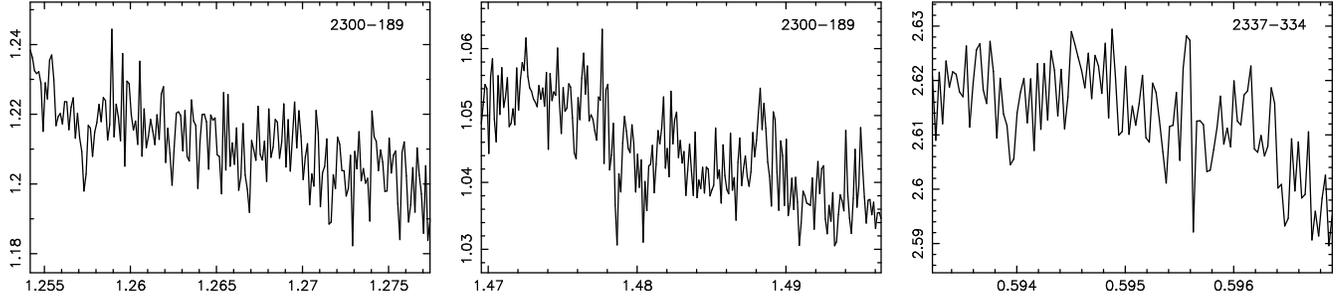
  %other layout saved in 0003inc.tex
\addtocounter{figure}{-1}
	\vspace{4.5cm} \setlength{\unitlength}{1in}
\includegraphics{2300-189-HI.ps}
\includegraphics{2300-189-OH.ps}
\includegraphics{2337-334.ps}
\caption{{\em Continued.} }
\end{figure*}

%\subsubsection{\HI\ and OH limits}
%\subsubsection{Measurements of column density}

In Table \ref{sum} we summarise the results of this phase of the
survey.  The column densities over the searched redshift ranges are
derived from the integrated optical depths measured from the spectra.
For \HI\ 21-cm this is obtained from
\begin{equation}
%N_{\rm HI}=\frac{1.823\times10^{18}.T_{\rm s}}{f}\left.\int\right.\tau dv,
N_{\rm HI}=1.823\times10^{18}.\frac{T_{\rm s}}{f}\left.\int\right.\tau dv
\label{e1}
\end{equation}
where $T_{\rm s}$ is the spin temperature of the gas (K). For OH in thermal equilibrium the column density is given by
%(e.g. \citealt{pkn+04})
\begin{equation}
N_{\rm OH}=X\times10^{14}.\frac{T_{\rm x}}{f}\left.\int \right.\tau dv,
\label{eqn:oh}
\end{equation}  % X factors are from \citet{hgb87}
where $T_{\rm x}$ is the excitation temperature and $X=1.61$, 2.38 and
4.30 for the 4751, 1667 and 1665 MHz lines, respectively.  In both
equations, $f$ is the covering factor of the background continuum flux
by the absorber and $\int\tau dv$ is the integrated optical depth of
the line (\kms), where $\tau=\sigma_{{\rm rms}}/S_{\rm cont}$ in the
optically thin ($\tau\lapp0.3$) regime.  
\begin{table*} 
\centering
\begin{minipage}{165mm}
\caption{Summary of the searches for centimetre absorption lines in
the hosts of red quasars and radio galaxies. $N$ is the column density
derived from the optical depth quoted in Table \ref{obs} ($3\sigma$
upper limits are given), $T_{\rm s}$ is the spin temperature of the
\HI~21-cm line, $T_{\rm x}$ is the excitation temperature of the OH
line and $f$ the respective covering factor. The column density of the
latter species has been calculated from the 1667 MHz line (Equation
\ref{eqn:oh}), except for 1535$+$004 which is from the 4751 MHz line.}
\begin{tabular}{@{}l c c c c c  @{}} 
\hline
Source & $z_{\rm host}$ & Transition &  $z$-range     & $N$ [\scm%~K$^{-1}$
                                                                 ]                          & Comments \\
\hline
%HAVE COMMENTED OUT THOSE YET TO BE REDONE
%%%%%...        & ...     & OH$_{1665}$  & 0.3242--0.3307  & $<7.2\times10^{13}.\,(T_{\rm x}/f)$ & \\
%%%%%...        & ...     & ...          & 0.3258--0.3323  & $<7.2\times10^{13}.\,(T_{\rm x}/f)$ & ...\\
0108$+$388 & 0.66847 & OH   &0.6624--0.6741  & $<2.0\times10^{14}.\,(T_{\rm x}/f)$ & $N_{\rm HI}=8.1\times10^{19}.\,(T_{\rm s}/f)$~\scm\ at $z=0.66847$$^a$\\
%%%%%...        & ...     & OH$_{1665}$  & 0.6621--0.6729  & $<1.5\times10^{14}.\,(T_{\rm x}/f)$ & \\
0114$+$074 & 0.342   & OH   & 0.3382--0.3457  & $<4.3\times10^{13}.\,(T_{\rm x}/f)$ & \\
%%%%%...        & ...     & OH$_{1665}$  & 0.3377--0.3441  & $<1.0\times10^{14}.\,(T_{\rm x}/f)$ & \\
%0131$-$001 & 0.879   & \HI          & 0.8891--0.9043  & $<8.1\times10^{17}.\,(T_{\rm s}/f)$ & Severe RFI required splicing of band \\
%...        & ...     & ...          & 0.8833--0.8861  & $<1.3\times10^{18}.\,(T_{\rm s}/f)$ & \\
%...        & ...     & ...          & 0.8704--0.8803  & $<1.9\times10^{18}.\,(T_{\rm s}/f)$ & \\
%..        & ...     & ...          & 0.8626--0.8650  & $<1.9\times10^{18}.\,(T_{\rm s}/f)$ & \\
 0131$-$001       & 0.879   & OH   & 0.8588--0.8947  & $<3.1\times10^{13}.\,(T_{\rm x}/f)$ & No coverage over $z= 0.883-0.885$ due to RFI\\
0153$-$410 & 0.226   & OH   & 0.2099--0.2323  & $<3.3\times10^{14}.\,(T_{\rm x}/f)$ & \\
%%%%%...        & ...     & OH$_{1665}$  & 0.2086--0.2309  & $<1.4\times10^{14}.\,(T_{\rm x}/f)$ & \\
%%%0202$+$149 &0.405 & \HI &0.1400--0.1473 & $<2.7\times10^{17}.\,(T_{\rm s}/f)$ & 0114+074 calibration \\
%%%...& ...& OH  &0.3382--0.3468 & $<3.9\times10^{13}.\,(T_{\rm x}/f)$ & ...\\
%%%...& ...& OH$_{1665}$ &0.3366--0.3452& $<7.1\times10^{13}.\,(T_{\rm x}/f)$ & ...\\
0202$+$149 & 0.405   & OH   &0.4011--0.4094   & $<2.5\times10^{13}.\,(T_{\rm x}/f)$ &  \\
%%%%%...        & ...     & OH$_{1665}$  & 0.4007--0.4078  & $<6.4\times10^{13}.\,(T_{\rm x}/f)$ & \\
%%%%%0454+066 & 0.405 & \HI & 0.4036--0.4105 & $<2.1\times10^{18}.\,(T_{\rm s}/f)$ & \\
0500$+$019 & 0.58457 & OH   & 0.5789--0.5902  & $<5.6\times10^{13}.\,(T_{\rm x}/f)$ & $N_{\rm HI}=7.0\times10^{18}.\,(T_{\rm s}/f)$~\scm\ at $z=0.58457$$^a$\\
%%%%%...        & ...     & OH$_{1665}$  & 0.5771--0.5884  & $<8.5\times10^{13}.\,(T_{\rm x}/f)$ &\\
1107$-$187 & 0.497   & OH   & 0.4927--0.5021  & $<8.0\times10^{13}.\,(T_{\rm x}/f)$ & \\
%%%%%...        & ...     & OH$_{1665}$  & 0.4910--0.5004  & $<2.1\times10^{14}.\,(T_{\rm x}/f)$ & \\
1353$-$341 & 0.2227  & OH   & 0.2047--0.2278 & $<3.3\times10^{14}.\,(T_{\rm x}/f)$ &\\
%%%%%...        & ...     & OH$_{1665}$  & 0.2016--0.2264  & $<2.0\times10^{14}.\,(T_{\rm x}/f)$ & \\
1355$+$441 & 0.6451  & OH   &0.6395--0.6509 & $<3.8\times10^{13}.\,(T_{\rm x}/f)$ & $N_{\rm HI}=3.3\times10^{19}.\,(T_{\rm s}/f)$~\scm\ at $z=0.6451$$^b$\\
%%%%%...        & ...     & OH$_{1665}$  & 0.6360--0.6473  & $<6.8\times10^{13}.\,(T_{\rm x}/f)$ &\\
1450$-$338 & 0.368   & \HI          & 0.3632--0.3710  & $<6.6\times10^{17}.\,(T_{\rm s}/f)$ &\\
...        & ...     & OH   & 0.3468--0.3757  & $<1.9\times10^{14}.\,(T_{\rm x}/f)$ &\\
%%%%%...        & ...     & OH$_{1665}$  & 0.3420--0.3730  & $<2.4\times10^{14}.\,(T_{\rm x}/f)$ &\\
1535$+$004 & 3.497   & \HI & 3.4864--3.5092  & $<1.7\times10^{18}.\,(T_{\rm s}/f)$ & \\
...& ...& OH &3.4818--3.5115 & $<9.2\times10^{13}.\,(T_{\rm x}/f)$& \\
1555$-$140 & 0.097   & \HI          & 0.0971          & $4.2\times10^{19}.\,(T_{\rm s}/f)$  &\\
%2252--090 &  0.6064 & OH  & 0.6002--0.6125& --& \\
%...& ...& OH$_{1665}$&0.5983--0.6106& --& \\
2300$-$189 & 0.129   & \HI  & 0.1123--0.1318  & $<1.6\times10^{18}.\,(T_{\rm s}/f)$ & \\
..        & ...     & OH   & 0.1145--0.1343  & $<1.2\times10^{14}.\,(T_{\rm x}/f)$ & $N_{\rm OH}\approx4\times10^{14}.\,(T_{\rm x}/f)$~\scm\ at $z=0.1263$$^c$? \\
%%%%%...        & ...     & OH$_{1665}$  & 0.1132--0.1337  & $<8.5\times10^{13}.\,(T_{\rm x}/f)$ &\\
2337$-$334 & 1.802   & OH   & 1.7929--1.8117  & $<4.2\times10^{13}.\,(T_{\rm x}/f)$ & \\
%%%%%...        & ...     & OH$_{1665}$  & 1.7896--1.8084  & $<6.9\times10^{13}.\,(T_{\rm x}/f)$ & \\
\hline 
\end{tabular}
%{\flushleft References: $^a$\protect\citet{cmr+98}  $^b$\protect\citet{kb03}.}
{\flushleft Notes: $^a$From \protect\citet{cmr+98},  $^b$\protect\citet{vpt+03}, $^c$see Sect. \ref{2300}.}
\label{sum}
\end{minipage}
\end{table*}

\subsubsection{\HI\ absorption in a radio galaxy at $z\sim0.10$}
\label{1555}

\begin{figure}
%\vspace{6.6cm}
%\special{psfile= 1555-dfreq.ps hoffset=-23 voffset=220 hscale=38 vscale=38 angle=270}
\vspace{5.2cm}
%\special{psfile=det1.ps hoffset=-130 voffset=270 hscale=62 vscale=65 angle=270}
\includegraphics{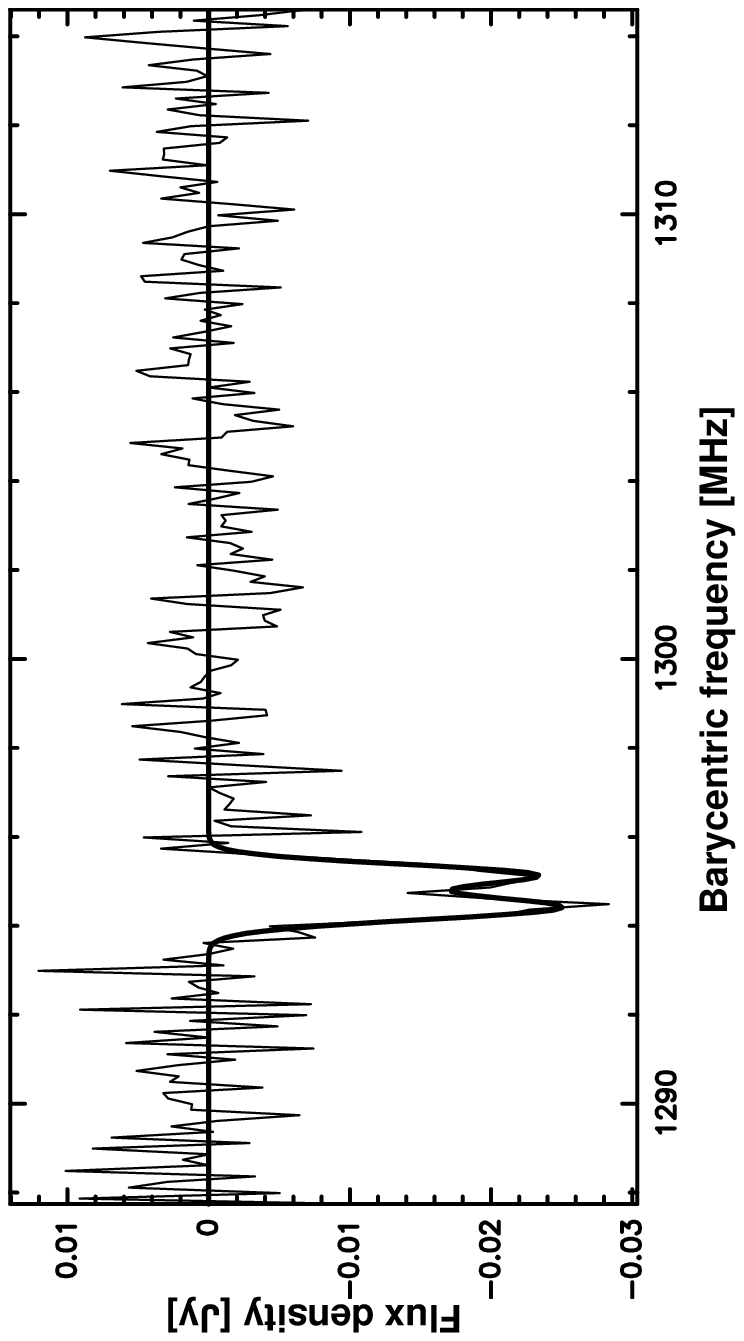}

\caption{\HI~21-cm absorption in the radio galaxy PKS 1555--140. In
this and Fig. \ref{2300-spectrum}, the black line line shows a
two-component Gaussian fitted to the residual spectrum after
subtracting a first order baseline from the line-free channels.}
\label{spectrum}
\end{figure}

In Fig. \ref{spectrum} we show the one clear detection obtained during
this leg of the survey, that of \HI\ in the radio galaxy PKS
1555$-$140.  The velocity integrated optical depth of the line,
$\int\tau dv = 23\pm2$ \kms, gives an atomic hydrogen column density of $N_{\rm
HI}=4.2\pm0.4\times10^{19}.\,(T_{\rm s}/f)$ \scm\ which is at the
upper end of the range of the known redshifted 21-cm absorbers (Table
\ref{t2}) and for the canonical $T_{\rm s}/f\gapp100$~K, this implies
a neutral hydrogen column density in excess of $N_{\rm
HI}\gapp10^{21}$ \scm. The central peak in the absorption profile
occurs at a redshift of $z_{\rm abs}=0.0971\pm0.0001$, where the
uncertainty quoted is a single channel width. This compares well with
the redshift of $z=0.097$ for this galaxy. We have recently undertaken
optical observations of this radio galaxy and some of the several
smaller galaxies surrounding it, in order to further study this
interacting system. Our results will be presented in Whiting et
al. (2006, in prep.).

\subsubsection{Possible OH absorption in a radio galaxy at $z\sim0.13$}
\label{2300}

\begin{figure}
\vspace{5.2cm}
%\special{psfile=det2.ps hoffset=-130 voffset=270 hscale=62 vscale=65 angle=270}
\includegraphics{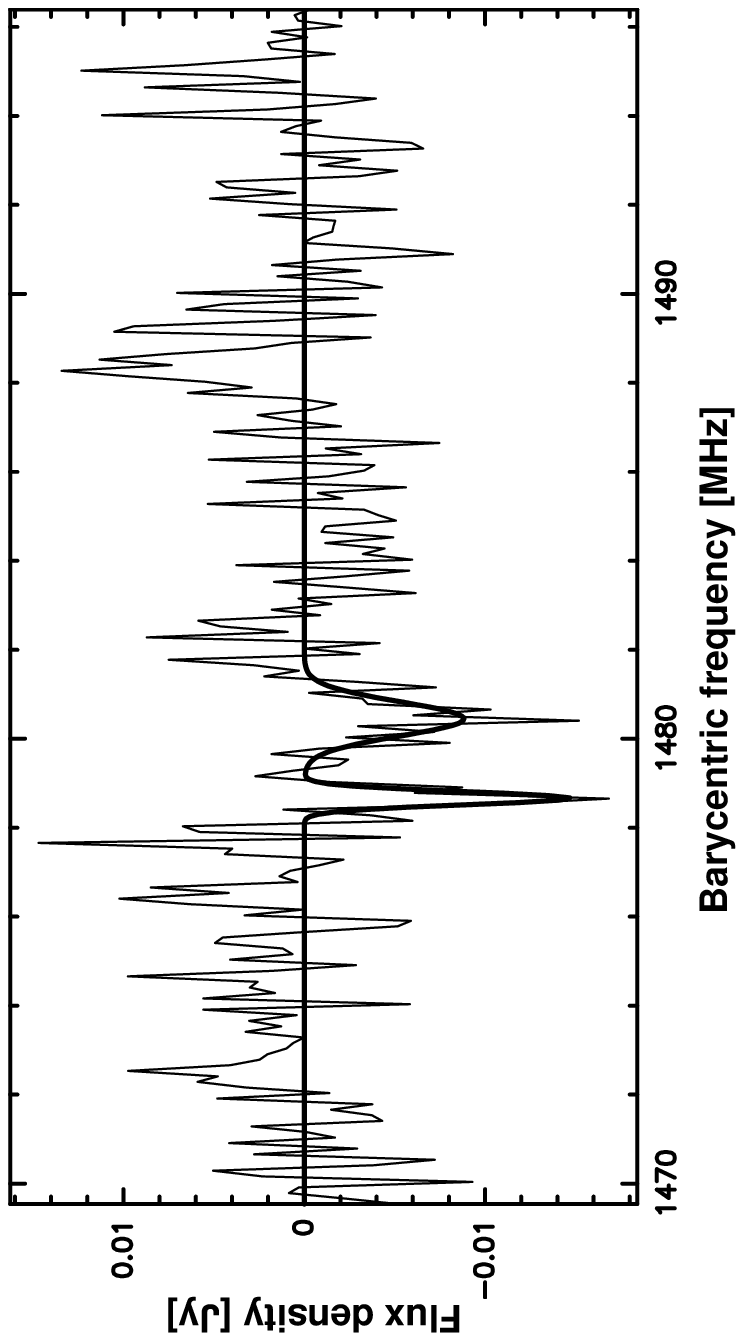}
\caption{The possible OH absorption feature in PKS 2300--189. }
\label{2300-spectrum}
\end{figure}

In the $\approx1.48$ GHz spectrum of 2300--189 (Fig. \ref{spectra})
there are two troughs that appear to be separated by $\approx2$~MHz,
as would be expected from the Lambda-doubled lines of OH in the ground
state. The statistical significance of these lines is marginal,
although their peaks are the only points at $>3\sigma$ away from the
first-order baseline fit indicated in the figure\footnote{Note that
here $\sigma=4.9$~mJy, which is calculated from the
baseline-subtracted spectrum (Fig.  \ref{2300-spectrum}). Thus, unlike
the $\sigma_{{\rm rms}}$ value quoted in Table \ref{obs}, it does not
include the large-scale gradient seen in the spectrum in
Fig. \ref{spectra}.}. A one-sample runs test shows that there are
significant deviations (at the 99.5\% level) from randomness in the
baseline-subtracted spectrum, and if one breaks the spectrum into four
quarters, the only part that shows significant deviations is the
quarter containing these two putative lines (then at the 99\% level).

From Gaussian fits to the spectrum (Fig. \ref{2300-spectrum}),
we find that the two features occur at $1478.66\pm0.05$ and
$1480.44\pm0.12$ MHz.
%\footnote{Derived from the offsets of $877\pm10$ and $517\pm24$ \kms\ from the centre observed frequency.}.  
If these are due to absorption by OH in the ground state, then the
lower frequency gives a redshift of $z=0.12629\pm0.00004$ for the 1665
MHz line and the higher frequency $z=0.12626\pm0.00009$ for the 1667
MHz line. These are in excellent agreement and suggest foreground molecular
absorption at an offset of $\approx-700$ \kms\ from the background
radio source.

From the Gaussian fits, we calculate integrated optical depths of
$\int\tau dv = 1.05\pm0.32$ and $1.63\pm0.44$ \kms\ for the 1665 and
1667 MHz lines, respectively. These give total OH column densities of
$N_{\rm OH} = 4.5\pm1.4\times10^{14}.\,(T_{\rm x}/f)$ and
$3.9\pm1.1\times10^{14}.\,(T_{\rm x}/f)$ \scm, respectively, thus
being consistent with the ratio of line strengths expected for OH in
thermal equilibrium.  We also searched for \HI\ in this source
(Fig. \ref{spectra}), and although there may be a feature at
$\approx1257$ MHz (corresponding to $z\approx0.129$, the redshift of
the galaxy), no \HI\ absorption was detected to a $3\sigma$ limit of $N_{\rm
HI}>1.6\times10^{18}.\,(T_{\rm s}/f)$ \scm\ per channel at
$z\approx0.1263$ (1261 MHz).

The width of the 1667 MHz line is a typical \citep{cdn99,kc02a}
$180\pm60$ \kms.  If we assume that the \HI\ has the same width, the
limit becomes $N_{\rm HI}<3.9\times10^{18}.\,(T_{\rm s}/f)$ \scm\ for
data ``smoothed'' to a single channel. This gives a ratio of
$\frac{N_{\rm OH}}{N_{\rm
HI}}\gapp10^{-4}\,\frac{T_x}{T_s}.\frac{f_{\rm HI}}{f_{\rm OH}}$.
%Forthe canonical $T_x/f = 10$ K and $N_{\rm H_2}=10^7\times N_{\rm OH}$
%(e.g. \citealt{kc02a}), this implies a molecular hydrogen column
%density of $N_{\rm H_2}\gapp10^{22}$ \scm. 
Clearly, however, such a weak ``detection'' requires confirmation
before we can confidently speculate on the molecular fraction and
other related properties.

%%%%%\subsubsection{Possible absorption towards a radio galaxy at $z\sim0.2$}
%%%%%\label{1353-341}

%%%%%\begin{figure}
%%%%%\vspace{13.5cm}
%%%%%\special{psfile=1353-341XX.log.ps hoffset=-23 voffset=420 hscale=65 vscale=65 angle=270}
%%%%%\special{psfile=1353-341YY.log.ps hoffset=-23 voffset=230 hscale=65 vscale=65 angle=270}
%%%%%\caption{OH absorption in the radio galaxy PKS 1353--341. Top: XX polarisation. Bottom:%%%%% YY polarisation.}
%%%%%\label{spectrum}
%%%%%\end{figure}
%%%%%However, as well as the different strengths in each polarisation
%%%%%{\bf If kept, well have to\\add to plots - different symbol\\discuss in later sections\\add to abstract}

\subsubsection{\HI\ and OH limits}

Apart from the detection in 1555$-$140, useful \HI\ data were obtained
for a further three sources, all of which reach limits of $N_{\rm
HI}\sim10^{18}.\,(T_{\rm s}/f)$ \scm\ (Table \ref{sum}), comparable
with many of the detections (Table \ref{t2}). Of these, \citet{vpt+03}
found \HI\ in the hosts of 19 of the 57 sources for which useful data
were obtained in a survey of redshifted compact radio sources. A
detection rate of $\approx30\%$ for \HI\ in near-by active galactic
nuclei (AGN) was also found by \citet{vke+89}, perhaps due to cold gas
being located in the dusty environment encircling the nucleus (see
\citealt{mot+01} and Sect. \ref{sec-phot}).

\citet{cmr+98} also searched for \HI\ absorption, in the hosts of
reddened quasars, and obtained a detection rate of $80\%$, cf.
$\approx10\%$ for Mg{\sc \,ii}-selected objects (generally low
redshift DLAs, e.g. \citealt{rt00}). However, the red quasars consist
of a sample of just five and the detection rate in the Mg{\sc
\,ii}-selected objects is based upon the number of systems expected to
have $N_{\rm HI}\geq2\times10^{20}$ cm$^{-2}$ in optically selected
samples. For DLAs occulting radio-loud background continua, the
detection rate of \HI\ 21-cm is $\approx50$\% for the published
sources (see \citealt{cmp+03}), most of which were originally
identified through the Mg{\sc \,ii} lines \citep{cw06}. These have
neutral hydrogen column densities in the range of $N_{\rm HI}=0.02 -
6\times10^{19}.\,(T_{\rm s}/f)$ \scm, whereas the red
quasars of \citet{cmr+98} have values of $0.8 -
8\times10^{19}.\,(T_{\rm s}/f)$ \scm, a range of higher values which
nevertheless overlaps considerably with that of the DLAs. In this work
we have searched two of the Carilli et al.  sources (0108+388 \&
0500+019) for OH absorption (discussed in Sect. \ref{known}). The
remaining \HI\ detections were towards 0218+357 and 1504+377, two of the
currently known molecular absorbers (Sect. \ref{other}).

\begin{figure}
\vspace{8.0 cm} \setlength{\unitlength}{1in} 
\begin{picture}(0,0)
\put(-0.2,3.7){\includegraphics{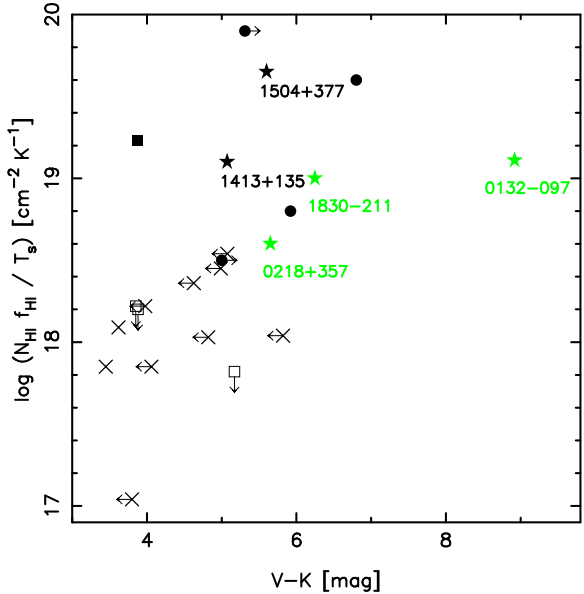}}
\end{picture}
\caption{The scaled \HI\ velocity integrated optical depth versus
optical--near-IR colour (where available) for the associated
absorbers. In this and the following figures, the unfilled symbols
designate $3\sigma$ limits. The squares represent our results and the
crosses the detections of \protect\citet{vpt+03}, where the
upper limits in $V-K$ designate $B-K$ values and the lower limits
$R-K$ values. The circles represent the other published detections
(Table \protect\ref{t2})
%where the colour for 0108+388 is a lower limit from $R-K=5.31$ (Table \ref{obs})
and the stars the four known OH absorbers (Table \ref{t3}); black
for the associated systems and coloured/grey for the intervening
absorbers.}
\label{f4}
\end{figure}

In order to investigate any possible correlation of hydrogen column
density with reddening, in Fig. \ref{f4} we show the scaled velocity
integrated optical depth of the \HI\
line\footnote{$1.823\times10^{18}.\left.\int\right.\tau dv$ which
gives $N_{\rm HI}.\,f{_{\rm HI}}/T_{\rm s}$.} versus the
optical--near-IR colour for our targets together with the previous
detections of associated \HI\ absorption (Table \ref{t2}).
From this we see a slight trend for the \HI\ line strength to increase with $V-K$
colour, although the spread is broad, particularly at large $V-K$
colours. Also, one of our targets has $V-K\geq5$, yet \HI\
remains undetected in the host at $N_{\rm HI}\sim10^{18}.\,(T_{\rm
s}/f)$ \scm.
\begin{table*}
\centering
\begin{minipage}{107mm}
\caption{The known redshifted ($z_{\rm abs}\gapp0.1$) \HI~21-cm
absorbers. Absorber types are: BLRG--broad line radio galaxy,
CSS--compact steep spectrum source, DLA--damped Lyman alpha absorption
system, GPS--gigahertz peaked spectrum source, Lens--gravitational
lens, OHM--OH megamaser, Red--red quasar, RG--radio galaxy. The number of each type is
given as well as the absorption redshift and column density ranges.
\label{t2}}
\begin{tabular}{@{}l c c c c @{}} 
\hline
Reference      & Type    & No.& $z_{\rm abs}$ & $N_{\rm HI}$ [\scm% ~K$^{-1}$
                                                                  ]\\
\hline
\multicolumn{5}{c}{ASSOCIATED ABSORBERS}\\
\hline
\citet{ubc91}   & RG     & 1  & 3.40       & $3\times10^{18}.\,(T_{\rm s}/f)$ \\
\citet{cps92}  & Red     & 1  & 0.25       & $1\times10^{19}.\,(T_{\rm s}/f)$ \\
\citet{cmr+98} & Red     & 3  & 0.58--0.67 & $0.8 - 8\times10^{19}.\,(T_{\rm s}/f)$ \\
\citet{mcm98}            & Lens & 1  & 2.63 & $8\times10^{18}.\,(T_{\rm s}/f)$ \\
\citet{ida03}  & CSS/Red & 1  & 1.19       & $4\times10^{19}.\,(T_{\rm s}/f)$ \\
\citet{vpt+03} & BLRG    & 1  & 0.22       & $7\times10^{17}.\,(T_{\rm s}/f)$\\
...            & CSS     & 7  & 0.19--0.80 & $0.1-2\times10^{18}.\,(T_{\rm s}/f)$\\
...            & GPS     & 10 & 0.08--0.65 & $0.07-3\times10^{19}.\,(T_{\rm s}/f)$ \\
...            & RG      & 1  & 0.24       & $1\times10^{18}.\,(T_{\rm s}/f)$ \\
\citet{pbdk05} & OHM     & 1  & 0.22       & $6 \times10^{18}.\,(T_{\rm s}/f)$ \\
This paper     & RG      & 1  & 0.10       & $4\times10^{19}.\,(T_{\rm s}/f)$\\
\hline
\multicolumn{5}{c}{INTERMEDIATE ABSORBERS}\\
\hline
\citet{cry93}            & Lens & 1  & 0.69 & $1\times10^{19}.\,(T_{\rm s}/f)$ \\
\citet{lrj+96}$^*$       & Lens & 1  & 0.19 & $\approx2\times10^{18}.\,(T_{\rm s}/f)$ \\

\citet{cdn99}            & Lens & 1  & 0.89 & $1\times10^{19}.\,(T_{\rm s}/f)$ \\
\citet{kb03}             & Lens & 1  & 0.76 & $1\times10^{19}.\,(T_{\rm s}/f)$ \\
\citet{kc02}$^{\dagger}$ & DLA  & 15 & 0.09--2.04 &  $0.02-6\times10^{19}.\,(T_{\rm s}/f)$ \\ 
\citet{dgh+04}           & DLA  & 1  & 0.78 & $2\times10^{19}.\,(T_{\rm s}/f)$ \\
\citet{kse+06}           & DLA  & 1  & 2.347 & $4\times10 ^{17}.\,(T_{\rm s}/f)$ \\ 
Zwaan et al. (in prep.) & DLA & 2 & $\sim0.6$ & --\\ 
\hline
\end{tabular}
{Notes: $^*$A possible second lensing system towards 1830--211 (see
Sect. \ref{other}). $^{\dagger}$See the paper for the full reference
list and \protect\citet{cmp+03} for the calculated column densities. 
Note that since PKS 1413+135 is an associated system, it has
been included in the top panel.}
\end{minipage}
\end{table*}

%\newpage

\subsection{Comparison with known redshifted OH absorbers}
\label{known}

\subsubsection{Colours}
\label{colours}

\begin{table*}
\centering
\begin{minipage}{133mm}
\caption{Redshifted OH absorbers towards AGN and QSOs. $N_{\rm OH}$ is
the column density from the OH 1667 MHz line. Those in the top panel
constitute the known millimetre absorption systems. Note that all
sources also feature in Table \ref{t2}, in which 2 are associated and
3 are intermediate \HI\ absorbers.
\label{t3}} 
\begin{tabular}{@{}l c c c c c c c c @{}} 
\hline
Source     & $z_{\rm abs}$ 
                     & $z_{\rm em}$ 
                             & $N_{\rm HI}$ [\scm% ~K$^{-1}$
                                                 ]                & Ref. & $N_{\rm OH}$ [\scm% ~K$^{-1}$
                                                                                             ]             & Ref.  & $V-K$ & Ref.\\
\hline
0218$+$357 & 0.68466 & 0.94  & $4\times10^{18}.\,(T_{\rm s}/f)$   & 1 & $1.1\times10^{14}.\,(T_{\rm x}/f)$ & 6,7   & 6.167 & 9,10\\
1413$+$135 & 0.24671 & 0.247 & $1.3\times10^{19}.\,(T_{\rm s}/f)$ & 2 & $5.5\times10^{12}.\,(T_{\rm x}/f)$ & 6     & 5.072 & 11,10\\ 
1504$+$377 & 0.67335 & 0.673 & $4.5\times10^{19}.\,(T_{\rm s}/f)$ & 3 & $1.1\times10^{14}.\,(T_{\rm x}/f)$ & 6     & 5.348 & 12,13\\
1830$-$211 & 0.88582 & 2.507 & $1\times10^{19}.\,(T_{\rm s}/f)$   & 4 & $3.5\times10^{14}.\,(T_{\rm x}/f)$ & 4     & 6.246 & 9,10\\
\hline
0132$-$097 & 0.764 & 2.2 & $1.3\times10^{19}.\,(T_{\rm s}/f)$   & 5 & 0.35 to $3.5\times10^{16}$& 8 & 8.92 & 14\\
\hline
\end{tabular}
{References: (1) \citet{cry93}, (2) \citet{cps92}, (3) \citet{cmr+98},
(4) \citet{cdn99}, (5) \citet{kb03}, (6) \citet{kc02a}, (7)
\citet{kcdn03}, (8) \citet{kcl+05}, (9) \citet{hmr+01}, (10) 2MASS,
(11) \citet{hb89}, (12) SDSS, \citet{aaa+06}, (13) \citet{srkr96}, (14) \citet{glw+02}.}
\end{minipage}
\end{table*}

\begin{figure}
\vspace{8.0 cm} \setlength{\unitlength}{1in}
\begin{picture}(0,0)
\put(-0.2,3.7){\includegraphics{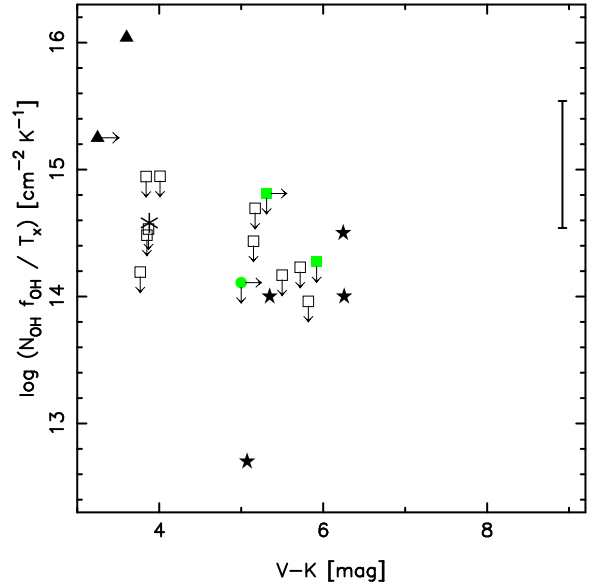}}
\end{picture}
\caption{The scaled OH velocity integrated optical depth versus
colour. The shapes are as per Fig. \ref{f4} with the coloured/grey
markers representing systems where \HI\ 21-cm absorption has been
detected and the unfilled markers representing the blind OH searches,
i.e. no other absorption known at the observed (host) redshift. The
triangles represent the two OH absorbing ULIRGs and the asterisk shows
the tentative detection towards 2300--189 (Sect. \ref{2300}). The
error bar shows the OH column density range of 0.35 to
$3.5\times10^{16}$ \scm ~for 0132$-$097 \citep{kcl+05}, converted to
the integrated optical depth assuming the canonical excitation
temperature of 10 K and a covering factor of unity. Note that for our
non-detections, in this and the following plots, we multiply the
integrated optical depth per channel width (Table \ref{sum}) by
$\sqrt{200/\Delta v}$ in order to convert these to the limit in a
single channel ``smoothed'' to 200 \kms, the typical line-width of the
known OH absorbers.  }
\label{f3}
\end{figure}

The objective of our sample selection was to choose sources with large
optical--near-infrared colours, which could possibly be attributed to
an abundance of dust, and thence molecules, along the
line-of-sight. In Fig.~\ref{f3} we plot the scaled velocity integrated
optical depth\footnote{$2.38\times10^{14}.\left.\int\right.\tau dv$
(1.61 for 1535$+$004) giving $N_{\rm OH}.\,f{_{\rm OH}}/T_{\rm x}$.} as a
function of the $V-K$ colour for our observations and the five known
OH absorbers. In addition, we include 0758+143 from \citet{ida03} and
0902+343 from \protect\citet{cb03}, who detected \HI\ [$N_{\rm
HI}=3.4\times10^{18}.\,(T_{\rm s}/f)$ \scm] but not OH
[$N_{\rm OH}<6\times10^{13}.\,(T_{\rm x}/f)$ \scm\ at a
$3\sigma$ level].

In Fig. \ref{f3} we also show the two other known redshifted OH
absorption systems, in the ultra luminous infra-red galaxies (ULIRGS)
19154+2704 ($z=0.099$) and 12107+3157 ($z=0.207$)
(\citealt{dg00,dg02}). These sources, however, are different in nature
to the radio-loud AGN that make up our sample and the remaining five
known absorbers. In particular, the optical--near-infrared colour of
the ULIRGs measures just the starlight of a starbursting galaxy, which
we expect to be bluer than the typical host of a radio-loud AGN. The
physical interpretation of this colour is different to that of the
colour of radio-loud AGN, where it is a measure of the reddening of
the central nuclear source. We thus include these points in the plots
for the sake of completeness.

The large number of upper limits due to non-detections means that Fig. \ref{f3}
is of limited use for analysis purposes. There is, however, neglecting
the tentative detection towards 2300--189 (Sect. \ref{2300}), a clear trend
for the five sources with detected OH absorption to show increasing
line strength with increasing redness. We see that the limits imposed
by our non-detections are consistent with this trend, and so we may be
close to the sensitivity to detect OH in at least four of the
targets. However, since 0108+388 and 0500+019 are the only systems of
these in which \HI\ has been detected, any absorption which is
coincident with the reddening may not be located at the host galaxy
redshift. 

\begin{figure}
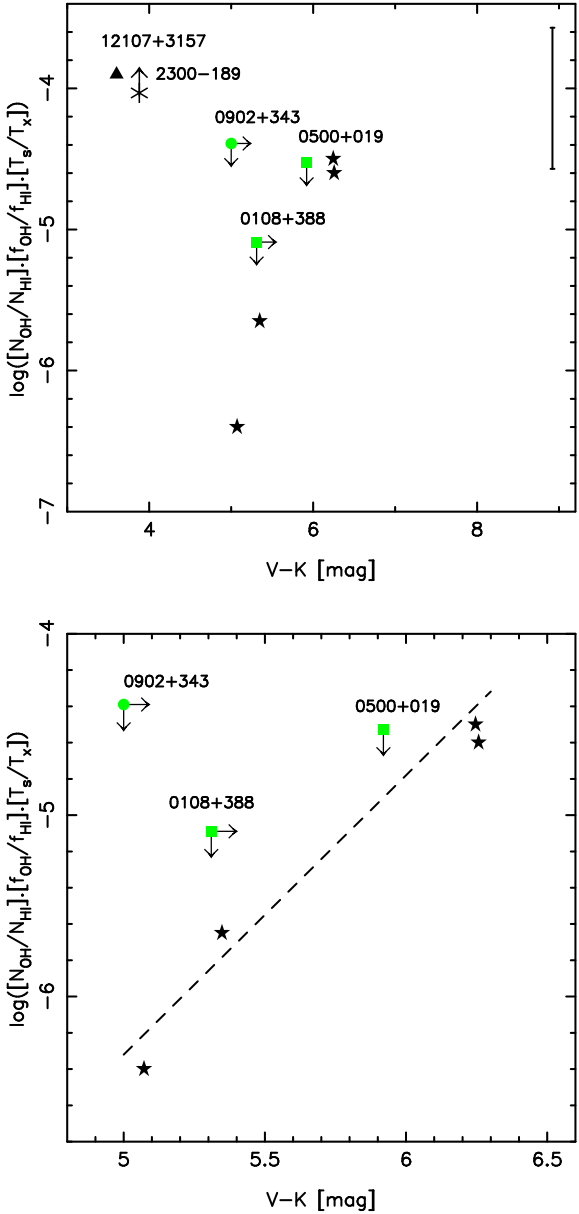

\vspace{16.4 cm} \setlength{\unitlength}{1in} 
\begin{picture}(0,0)
\put(-0.2,7.0){\includegraphics{OH-v-all.ps}} %from plots2/
\put(-0.2,3.7){\includegraphics{OH-v.ps}} %from plots2/
\end{picture}
\caption{The normalised OH line strength ($2.38\times10^{14}\int
\tau_{_{\rm  OH}}\, dv/1.82\times10^{18}\int\tau_{_{\rm HI}}\,
dv$) versus optical--near-IR colour. The bottom figure shows a detail
for the millimetre absorbers only, with the dashed line showing the
least-squares fit to these.}
\label{f2}
\end{figure}

In Fig. \ref{f2} we plot the OH velocity integrated optical depth normalised by that of the \HI\
against the optical--near-IR colour\footnote{For completeness we also
show 12107+3157 \citep{dg02} for which \HI\ 21-cm has also been
detected (J. Darling, priv.  comm.).}, where, as per
Figs. \ref{f4} and \ref{f3}, these are scaled in order to show
the ratio of $N_{\rm OH}. f_{_{\rm OH}}/T_x$ to $N_{\rm HI}. f_{_{\rm
HI}}/T_s$ on the ordinate. Again, neglecting the tentative detection
(Sect. \ref{2300}), the plot shows a significantly reduced scatter in
the correlation for the known OH absorbers (cf. Fig. \ref{f3}) and
provides good evidence that the reddening of these sources is due to
the presence of dust, be it in the intervening absorber (0132--097,
0218+357 and 1830--211), or located in the host (1413+135 and
1504+377). For the four sources also detected in millimetre-wave
transitions, the normalised OH velocity integrated optical
depth/optical--near-IR colour relationship can be characterised by
%\begin{equation}
\begin{multline} 
\frac{N_{\rm OH}}{N_{\rm HI}}\,.\frac{f_{\rm OH}}{f_{\rm
HI}}\,.\frac{T_{\rm s}}{T_{\rm x}} \equiv \frac{2.38\times10^{14}\int
\tau_{_{\rm OH}}\, dv}{1.82\times10^{18}\int\tau_{_{\rm HI}}\, dv}\\
= 10^{1.54(V-K) - 14.02},
\end{multline}  %with \usepackage{amsmath} allows line breaks
%\end{equation}
significant at
the 96\% level (Fig. \ref{f2}, bottom). The obvious
caveat in this conclusion is the fact that it is based on so few
detections, although Fig. \ref{f1} supports the steep increase of
molecular abundance with $V-K$ colour. We also note that 1830$-$211 has a low 
Galactic latitude, and a correspondingly high Galactic extinction 
($E(B-V)=0.46$, or $E(V-K)=1.33$. \citealt{sfd98}), which could contribute to the
observed reddening in this source. Since the resolution of the DIRBE maps is two 
orders of magnitude coarser than the size of the host galaxy, and there is limited
removal of contaminating sources at low latitudes, there is large uncertainty in the 
precise value of the extinction towards this source. Therefore, we do not
correct for this effect in our analysis, but simply note it here.

\subsubsection{Other factors}
\label{other}

We have selected our targets on the basis of the optical colours of
the four known redshifted OH absorption systems. According to
Fig. \ref{f3}, at least four of our targets could be red
enough, although only one of these is currently known to exhibit \HI\
absorption at the redshift observed. As stated previously, only two of
the five known systems exhibit the absorption in the host and so it is
perhaps worth summarising how these systems were originally
identified:
\begin{itemize}
\item CO $1\rr0$ absorption was detected in the host of BL Lac
  object PKS 1413+135 by \citet{wc94}, following the detection of a
  high column of \HI\ absorption at the same redshift
  \citet{cps92}. OH absorption was subsequently discovered by
  \citet{kc02a}.
\item Various millimetre-wave absorption lines were detected in a
  gravitational lens candidate at $z_{\rm abs}=0.685$ towards the
  BL Lac object B0218+357 \citet{wc95}. The redshift of the lens was previously
  determined from absorption features in the optical spectra of
  \citet{obs+92} and \citet{bpww93}, from which \citet{cry93} subsequently
  detected \HI\ absorption. OH absorption was detected by \citet{kc02a}.
\item Several millimetre-wave absorption lines were detected in
  the host of B3 1504+377 by \citet{wc96b}. This is an edge-on
  AGN, which would suggest high dust
  column densities along the line-of-sight to the nuclear continuum source.
  Optical lines were previously identified by \citet{sk94}, with \HI\ 
  and OH absorption found subsequent to the millimetre lines (\citealt{cmr+98}
  and \citealt{kc02a}, respectively).
\item Although the presence of a gravitational lens along the
  line-of-sight to the quasar PKS 1830--211 was known \citep{snps90},
  its redshift was not determined until \citet{wc96} identified
  several millimetre-wave absorption lines through a 14 GHz wide
  spectral scan of the 3-mm band\footnote{This is a technique we also
  tried towards visually obscured objects, although to no avail
  \citep{mcw02,cwmk03}.}.  Following this, \HI\ and OH absorption at
  this redshift were found by \citet{cdn99}. Note also, that an \HI\
  absorption feature has also been detected at $z_{\rm abs}=0.1926$
  towards the quasar \citep{lrj+96}. Since this has only $\approx20\%$
  the line strength of the other feature (Table \ref{t3}), summing
  these to give a total \HI\ column density towards the background
  quasar makes very little difference to Figs. \ref{f4} and
  \ref{f2}. Furthermore, molecular absorption was undetected at
  $z_{\rm abs}=0.1926$ \citep{wc98} and \citet{cmkl02} suggest that it
  is the $z_{\rm abs}=0.8858$ absorber which is responsible for the
  obscuration of the optical light.

\item After a long lull in the detection of new molecular absorbers in
rotational transitions at high redshift, \citet{kcl+05} detected OH
absorption in the $z_{\rm abs}=0.7645$ gravitational lens towards the
$z\sim2.2$ quasar PKS 0132--097 (PMN J0134--0931).  This was a known
gravitational lens in which the redshift was determined by Ca{\sc
\,ii} absorption lines \citep{hry+02}.  Following this, \HI\
absorption was detected by \citet{kb03} and with its extremely large
$V-K$ colour \citep{glw+02}, this was a prime target in which to
search for OH absorption\footnote{Indeed, we attempted the first
detection of OH absorption in this lens over both of the GMRT runs and
the WSRT run described here (Sect. \ref{observations}). Unfortunately,  on all
occasions severe RFI due to a mobile phone band operating at close to
945 MHz prevented even good data of the bandpass calibrator.}. Unlike
the other four molecular absorbers where HCO$^+$ is optically thick,
in this source HCO$^+$ remains undetected to $\tau<0.07$ at a
$3\sigma$ level \citep{kcl+05}.
%\footnote{We also failed to detect theHCO$^+$ $2\rr1$ transition with the ATCA in August 2005.}. 
This is unexpected since this optical depth limit gives a column
density of $N_{\rm HCO+}<2\times10^{11}$ \scm ~per channel ($T_{\rm x}
= 10$ K), thus being clearly short of the ratio $N_{\rm
OH}\approx30\times N_{\rm HCO+}$, shown by the other four redshifted
absorbers \citep{kc02a} and Galactic molecular clouds \citep{ll96a}.
\end{itemize}
Unlike most of our targets, four of the five known molecular
absorbers all have previously determined redshifts, from
either intervening lenses or absorption in the hosts of active
galaxies. As mentioned in Sect. \ref{sec-phot}, for the type-2 objects,
unified schemes for AGN imply that our line-of-sight to the nucleus is
obscured by a high column density of dust and we may expect absorption
from gas associated with this. Of our targets
which are type-2 radio galaxies (Table \ref{obs}):
\begin{enumerate}
\item 0114+074 has yet to be searched for \HI\ absorption, although
  Fig. \ref{f3} suggests our sensitivity is $\approx2$ dex short of
  that required to detect OH.

\item RFI marred the observations of \HI\ absorption redshifted to
  1011 MHz towards 0454+066. OH has yet to be searched for, but with
  $V-K = 4.3$, we would expect this to remain undetected.

\item Again, 0153$-$410 has yet to be searched for \HI\ absorption and
  with $V-K = 4.0$, Fig. \ref{f3} would suggest that $N_{\rm
  OH}\sim10^{12}.\,(T_{\rm x}/f)$ \scm, %~K$^{-1}$, 
  or two orders of
  magnitude lower than the sensitivity of our observations.

\item As above, no \HI\ absorption has yet been searched for in
  1353$-$341 and with $V-K = 3.8$, Fig. \ref{f3} suggests that another
  three orders of magnitude increase in the sensitivity of our
  observations would be required to reach the required $N_{\rm
  OH}\sim10^{11}.\,(T_{\rm x}/f)$ \scm.%~K$^{-1}$.

\item \HI\ of a high column density was detected in 1555--140,
  although again with a relatively low degree of reddening ($V-K =
  3.85$), we do not expect to detect OH.
  
\item \HI\ in 2300--189 was undetected down to a $3\sigma$ limit of
  $N_{\rm HI}<1.6\times10^{18}.\,(T_{\rm s}/f)$ \scm, although
we may have a detection of OH, which is unexpected according to
the $N_{\rm OH}$/$V-K$ relation (Fig. \ref{f3}).

\end{enumerate}
Because the scales probed by the high resolution continuum images of
these radio sources ($\sim$ tens of mas,
\citealt{mnp+86,dww+93,bpf+96,
lov97,jrb+99,ffp+00,mtc01,bgp+02,fpm+03}) are typically much smaller
than the scales that can be probed by the optical observations used to
derive the colours (typically 0.5~arcsec at best, and often integrated
over the entire host galaxy), one cannot be certain that optical and
radio sources are coincident on mas scales. An example could be
2300--189, in which we may have detected OH despite its blue
colour. The optical counterpart to the radio source could be strongly
reddened, but have its redness diluted by the inclusion of starlight
from the host galaxy. Nevertheless, Fig.\ref{f2} is indicates a 96\%
correlation between the strength of the OH absorption and the
optical--near-infrared colour,

\citet{sr97} find an excess of Mg{\sc ii} absorbers in the spectra of
BL Lac objects, which they attribute to the possible presence of a
foreground galaxy causing gravitational micro-lensing of the
background source.  Targetting the redshift of the Mg{\sc ii} absorber
may then be a useful way to search for new absorption systems.  As
well as the radio galaxies searched here, we have undertaken a survey
of BL Lac and narrow-line objects with the Effelsberg 100-m
telescope. However, due to severe RFI in the UHF bands, we may have to
re-observe these sources elsewhere. Also, as part of this survey we
are currently undertaking deep integrations for molecular absorption
in gravitational lenses towards reddened quasars, and shall report our
results in a later paper.

%\subsubsection{Spectral Energy Distributions }
%\label{seds}

\begin{figure*}
\begin{minipage}{170mm}
\includegraphics{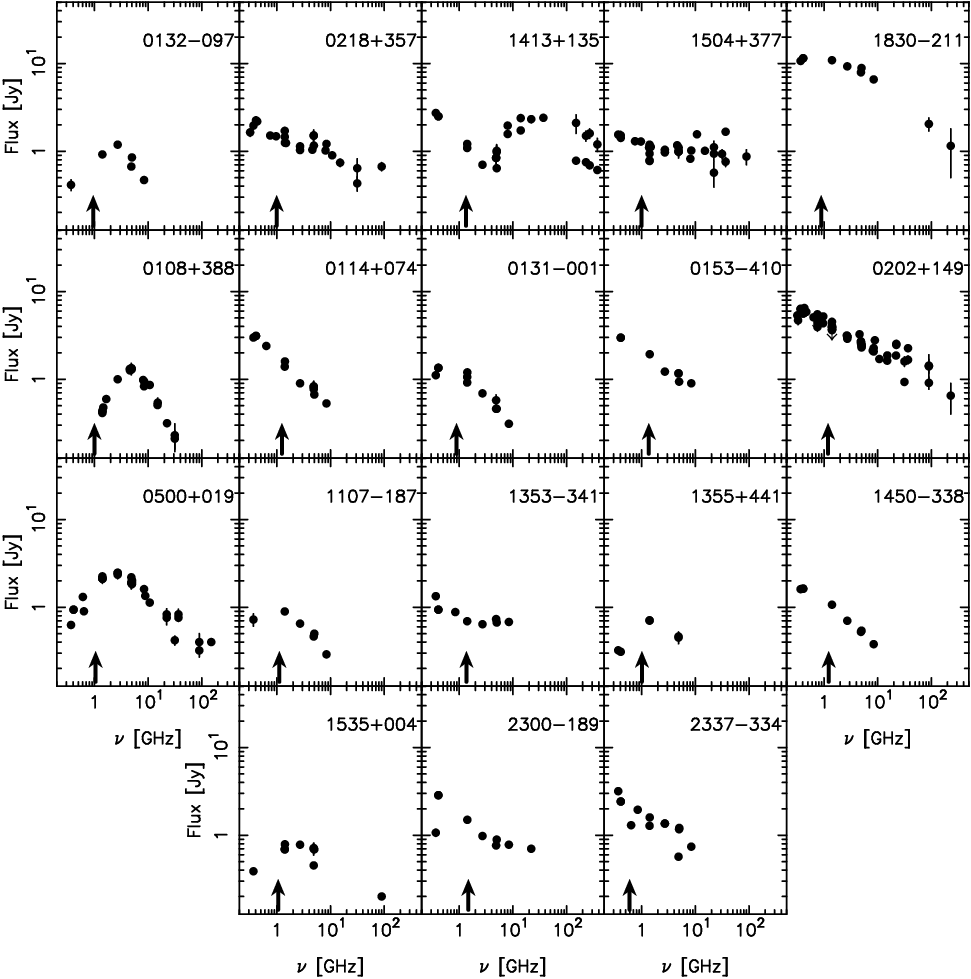}
\caption{The spectral energy distributions of the five known molecular
absorbers (top row) and the sources searched for OH absorption. All
flux densities have been obtained from the NASA/IPAC Extragalactic
Database (NED) (and references therein), as well as from
\citet{tvt+96} [1830$-$211, 0003$-$066 \& 0202$+$149], \citet{knk+99}
[0003$-$066 \& 2300$-$189] and \citet{cwmk03} [0500$+$019]. The arrows
indicate the frequency of the redshifted 1667 MHz OH line (4751 MHz
for 1535$+$004).}
\label{f-seds}
\end{minipage}
\end{figure*}

Finally for four of the five known systems we note that the molecular
absorption was initially identified via millimetre transitions, with
the OH line being confirmed afterward. Therefore, by definition, the
background continuum sources illuminating these absorbers all have
very flat radio spectra. Spectral indices of $\alpha>-0.2$ are
indicative of compact radio sources \citep{zpp84}, which perhaps
indicates that the coverage of the background continua by these
absorbers is large, thus maximising the observed optical
depth of the absorption line. For completeness, the spectral energy
distributions of the sources searched for OH are shown in Fig.~\ref{f-seds}.

\section{Summary}

We have presented the results from the first phase of our survey for
cosmologically redshifted atomic and molecular absorption lines in the
radio and microwave bands. All of our observations are towards quasars
and radio galaxies where we suspect, due to the reddening of these
objects, there may be large column of dust located somewhere along the
line-of-sight. In the absence of any previously known absorption
feature, we have observed at the host redshift in the hope of
detecting absorption due to cold dense gas located in the host galaxy.

Despite reaching limits comparable to the known \HI\ and OH
absorption systems, we only detect clear absorption in one source: \HI\
absorption in the $z=0.097$ radio galaxy PKS 1555--140. Besides this,
we only have complete searches for \HI\ in three other sources and
using the previously published detections of associated \HI\
absorption, we see only a slight trend for the \HI\ line strength
to increase with optical--near-IR colour.

Although we are searching at redshifts where no other absorption
features are yet known, we would expect, as for two of the five known
OH absorbers, molecular absorption within the host in at least some of
the 13 cases in which OH was searched. Presuming that this is not bad
luck due to the number of sources searched, when comparing our results to
the known redshifted molecular absorbers:
\begin{enumerate}
\item We find a steep increase in the molecular fraction of the
absorber with optical--near infrared colour of the quasar, as well as
a correlation of $\approx2\sigma$ significance between molecular line
strength (normalised by the atomic line strength) and the colour. If
reliable, this is a strong indicator that the reddening is due to dust
that has associated molecular gas, be this within an intervening
absorber or within the host galaxy.
  
\item Furthermore, at least four of the five known molecular
  absorption systems lie along the line-of-sight to quasars with
  extremely flat radio spectra. This suggests that the background
  sources are compact and that most of the observed flux is
  incident on the absorber, maximising the effective optical depth
  of the absorption line. Molecules in four of the five known
  systems were discovered through millimetre-wave observations,
  before being followed up with searches for OH in the centimetre
  band. In retrospect, the success of this method is not
  surprising, since the selection of sources with high millimetre
  fluxes introduces a bias in favour of more compact radio sources.

\item Finally, as well as being extremely red and occulting
  compact background continua, the known absorbers
  also have previously identified absorption features at the
  redshifts searched. The absorption systems comprise of three
  gravitational lenses in addition to an active galactic nucleus
  and BL Lac object, all of which we are currently searching as
  part of this survey. We shall report the results of these
  searches in subsequent papers.
\end{enumerate}

%\clearpage
\section*{Acknowledgments}

We wish to thank Annie Hughes who was the Duty Astronomer for the
remote ATCA observations, Bert Harms and Rene Vermeulen for the remote
WSRT observations, as well as the various telescope operators and
staff who greatly assisted our visits to the GMRT. We would also like
to thank Bob Sault, Alberto Bolatto, Peter Teuben and Tom Oosterloo
for their advice on the various intricacies of {\sc miriad}. We would
also like to thank the anonymous referee for their helpful in-depth
comments.

The Australia Telescope is funded by the Commonwealth
of Australia for operations as a National Facility managed by
CSIRO. The GMRT is run by the National Centre for Radio Astrophysics of
the Tata Institute of Fundamental Research. The Westerbork Synthesis
Radio Telescope is operated by the ASTRON (Netherlands Foundation for
Research in Astronomy) with support from the Netherlands Foundation
for Scientific Research NWO.

This research has made use of the NASA/IPAC Extragalactic
Database (NED) which is operated by the Jet Propulsion Laboratory,
California Institute of Technology, under contract with the National
Aeronautics and Space Administration. This research has made use of
NASA's Astrophysics Data System Bibliographic Services.

This publication makes use of data products from the Two Micron All
Sky Survey, which is a joint project of the University of
Massachusetts and the Infrared Processing and Analysis
Center/California Institute of Technology, funded by the National
Aeronautics and Space Administration and the National Science
Foundation.

%\bibliographystyle{apj} %same problem as before (m.tex)
%\bibliographystyle{mn2e}
%\bibliography{aa,ref}

%\bsp

\label{lastpage}

\end{document}